\documentclass[pra,reprint,longbibliography]{revtex4-1}
\usepackage{microtype}
\usepackage{amsmath,amsfonts,amssymb} 
\usepackage{physics}   

\usepackage{lmodern}
\usepackage{placeins}

\usepackage{graphicx}
\usepackage{color}
\usepackage{xfrac}
\usepackage{epstopdf}
\usepackage{blkarray} 
\usepackage{lipsum}
\usepackage{soul}

\usepackage{varioref}
\usepackage[colorlinks=true,
							pdfstartview=FitV, 
							linkcolor=blue,  
							citecolor=blue, 
							urlcolor=blue,
							linktocpage=true]{hyperref}
\usepackage{cleveref}
\usepackage{soul}

\usepackage[utf8]{inputenc}   
\usepackage{amsmath,braket,amssymb,graphicx,stackrel,color,hyperref}
\usepackage {graphicx}
\usepackage{float}

\usepackage{color}
\definecolor{darkblue}{rgb}{0.,0.,0.5}
\definecolor{darkred}{rgb}{0.5,0.,0.}
\definecolor{darkgreen}{rgb}{0.,0.5,0.}
\hypersetup{colorlinks=true, linkcolor=darkred, citecolor=darkblue, urlcolor=darkblue}

\newcommand{\HH}{\mathcal{H}}
\newcommand{\NN}{\mathcal{N}}
\renewcommand{\AA}{\mathcal{A}}

\renewcommand{\oc}{\omega_{\rm cyc}}
\newcommand{\lc}{\ell_{\rm cyc}}
\newcommand{\oz}{\omega_{\rm cav}}
\newcommand{\emath}{{\rm e}}

\renewcommand{\imath}{{\rm i}}
\renewcommand{\Re}[1]{\mathbb{R}{\rm e}\left[#1\right]}
\renewcommand{\Im}[1]{\mathbb{I}{\rm m}\left[#1\right]}

\begin{document}
	
\title{Vacuum-dressed cavity magnetotransport of a 2D electron gas}

\author{Nicola Bartolo}
\email{nicola.bartolo@univ-paris-diderot.fr}
\author{Cristiano Ciuti}
\email{cristiano.ciuti@univ-paris-diderot.fr}
\affiliation{Laboratoire Mat\'{e}riaux et Ph\'{e}nom\`{e}nes Quantiques, Universit\'{e} Paris Diderot, CNRS-UMR7162, 75013 Paris, France}

\begin{abstract}
We present a theory predicting how the linear  magnetotransport of a two-dimensional electron gas is modified by a passive electromagnetic cavity resonator where no real photons are injected nor created.
For a cavity photon mode with in-plane linear polarization, the dc bulk magnetoresistivity of the 2D electron gas is anisotropic.
In the regime of high filling factors of the Landau levels, the envelope of the Shubnikov-de Haas oscillations is profoundly modified and the resistivity can be increased or reduced depending on the system parameters. In the limit of low magnetic fields, the resistivity along the cavity-mode polarization direction is enhanced in the ultrastrong light-matter coupling regime.
Our work shows the crucial role of virtual polariton excitations in controlling the dc charge transport properties of cavity-embedded systems. 
\end{abstract}

\maketitle


\section{Introduction}

The physics of strong light-matter coupling has been attracting the interest of a large community thanks to the 
manipulation of  quantum states in cavity \cite{HarocheBook,Imamoglu2008} and circuit QED \cite{Wallraff2004,Scholkopf2008}, as well as for 
the control of linear and nonlinear optical properties in polaritonic systems \cite{RevModPhys.85.299}. 

During the last decade, there has been a considerable interest in the idea of using light to manipulate the electronic properties of materials, particularly for  light-induced
superconductivity \cite{Fausti189,Cavalleri2016}. This is a promising area of research because it can help to understand better the electronic properties of exotic materials
and give rise to new device functionalities.
More recently, a new frontier is opening up around the following general problem: is it possible to control the electronic transport properties of materials embedded in electromagnetic cavity resonators without injecting real photons? In other words, is it possible to have a vacuum-controlled transport of materials? Recently, experiments have suggested that electron transport can be modified due to the strong light-matter coupling in disordered molecular films embedded in metallic optical resonators \cite{Ebbesen2015,EbbesenPrivate} without shining light. 
A theoretical analysis of these complex systems has been based on simplified models describing a one-dimensional chain of two-level systems coupled to a cavity photon mode and studying exciton transport \cite{Pupillo2015,FeistPRL15} and charge conduction \cite{Hagenmuller2017}. Other works have studied cavity-mediated superconductivity \cite{Cavalleri2018}.The field is in its infancy and many questions remain open on the possibility of modifying the electronic transport via passive cavity resonators. 

A promising platform to explore and understand the effects of vacuum fields on electronic transport are cavity-embedded semiconductor 2D electron gas (2DEG) systems.
In presence of a perpendicular magnetic field, the inter-Landau-level cyclotron transition can be ultrastrongly coupled to a confined photon mode, meaning that the collective light-matter coupling can become comparable or larger than the confined photon mode and cyclotron frequencies \cite{HagenmullerPRB2010}, as demonstrated by several remarkable experiments \cite{ScalariScience2012,MaissenPRB2014,Kono2016,MaissenNJP2017,Huber2017,LiNatPhot18, Ultra_RMP, ParaviciniarXiv}.
The magnetotransport of a bare high-mobility 2DEG displays a rich phenomenology: for relatively-low magnetic fields, the bulk Drude-like longitudinal magnetoresistivity exhibits Shubnikov-de Haas oscillations \cite{Ando1982}; for high magnetic fields, the bulk becomes an insulator and the transport is dominated by the edge states with the emergence of the quantum Hall effects \cite{QuantumHallbook}. Recent experiments \cite{ParaviciniarXiv} have shown that the dc magnetotransport of a 2DEG can be significantly modified when the system is embedded in an electromagnetic cavity resonator.

In this article we present a theory revealing the cavity-controlled linear magnetotransport of a  2DEG in the dc regime where no real photon is injected nor created.
In Sec.~\ref{Sec:Model}, we present the model Hamiltonian, consider the current operators and 
determine the magnetoconductivity tensor via a linear response Kubo approach, consistently including the diamagnetic current contribution associated to the cavity mode, which is assumed to have an in-plane linear polarization. In Sec. ~\ref{results}, we discuss the main results. Finally, we draw our conclusions and future perspectives in Sec.~\ref{Sec:Conclusions}.
In Appendix~\ref{App:Details} we give some details of the calculations with some technical intermediate results.

\section{Theoretical framework}\label{Sec:Model}

\subsection{Hamiltonian and current operators}

\begin{figure}[t!]
	\includegraphics[width=0.9 \linewidth]{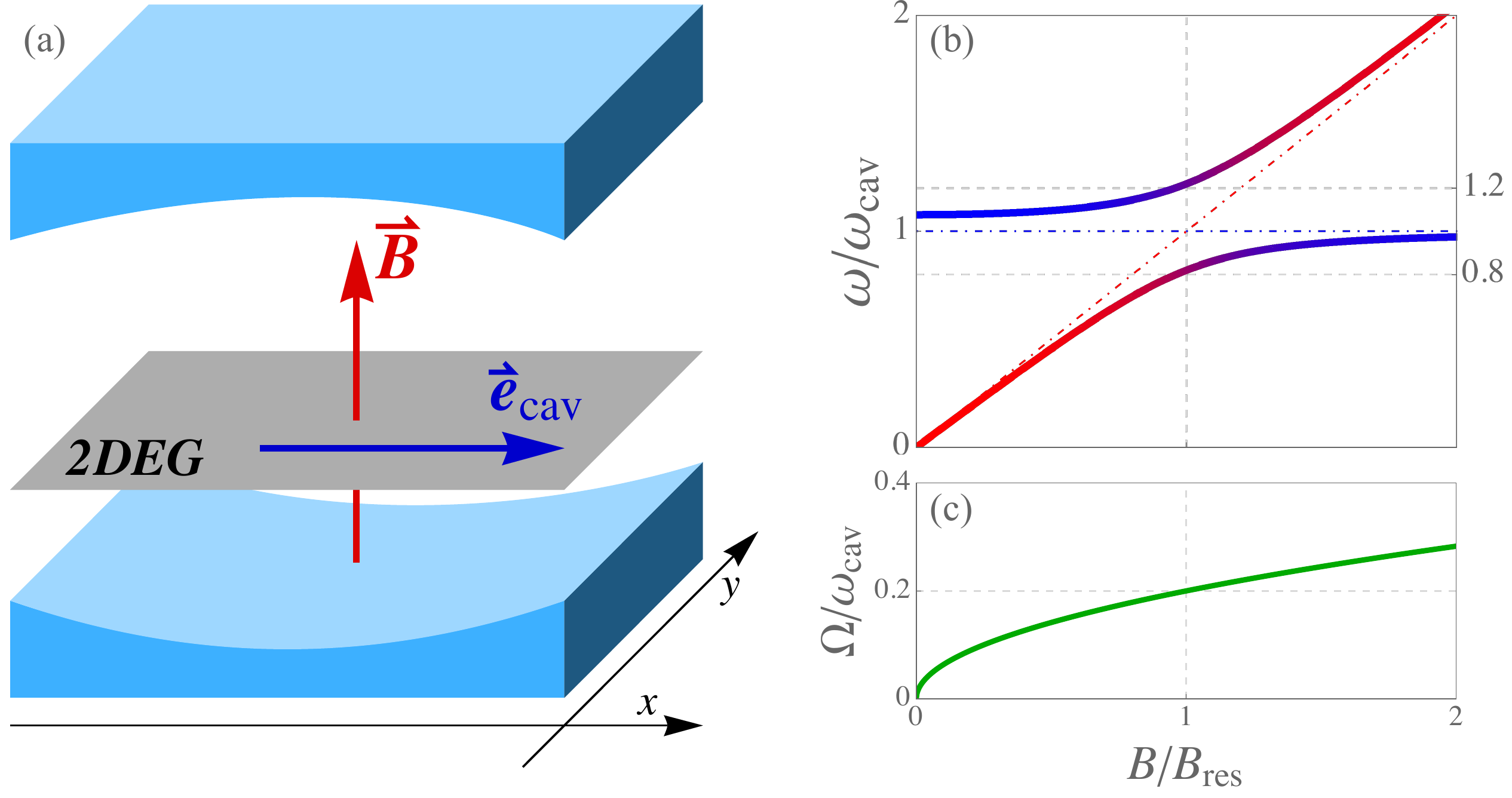}
	\caption{
		(a): sketch of a cavity-embedded 2D electron gas in presence of a perpendicular magnetic field $B$. 
		The cavity photon mode is assumed to have an electric field $\vec{e}_{\rm cav}$ along the $x$ direction.
		(b): cavity polariton frequencies as a function of the applied magnetic field.
		Dot-dashed lines are for the the bare cavity (blue) and cyclotron (red) frequencies.
		For $B=B_{\rm res}$ one has $\oc(B_{\rm res})=\oz$.
		(c): collective polariton Rabi frequency for $\Omega_{B=B_{\rm res}}=0.2\,\oz$.
		}
	\label{Fig:Scheme}
\end{figure}

Let us start by introducing the light-matter Hamiltonian describing a 2DEG coupled to a cavity mode in the presence of
a perpendicular magnetic field~$B$:
\begin{equation}
\label{Eq:2DEGHamiltonianRealSpace}
\hat{\HH}_{\rm lm}= \hbar\oz\,\hat{a}^\dag \hat{a} + \sum_i \frac{1}{2m_{\star}}\left(\hat{\bf p}_i+e\hat{\bf A}_i\right)^2,
\end{equation}
being $\hat{a}^\dag$ the creation operator of a cavity photon in the considered mode of frequency $\oz$, $e$ the electron charge and $m_{\star}$ the effective electron mass.
The sum runs over all the electrons, being $\hat{\bf p}_i$ the momentum operator for the $i$-th electron and
$\hat{\bf A}_i$ the electromagnetic vector potential operator at its position. We choose a gauge for the electromagnetic field with zero scalar potential and vector potential
\begin{equation}\label{Eq:VectorPotential}
\hat{\bf A}_i=\{A_0(\hat{a}+\hat{a}^\dag);B\hat{x}_i;0\},
\end{equation}
giving a static magnetic field $B$ perpendicular to the plane and an electric field operator $\hat{E}_{x,{\rm cav}}=\imath\,\oz A_0 \left(\hat{a}-\hat{a}^\dagger\right)$ \cite{CohenTannoudjiBook,MilonniBook}.
Hence, we are considering a spatially-uniform cavity mode polarized along $x$ in the region where the 2DEG is located. This is an excellent approximation for what experimentally achieved using metamaterial resonators \cite{ScalariScience2012,MaissenPRB2014,MaissenNJP2017,ParaviciniPRB17,ParaviciniarXiv}.
In the following, we will consider a 2DEG living on a rectangular area $\AA=L_x L_y$ with periodic boundary conditions along the $y$ direction.
A sketch of the system is presented in Fig.~\ref{Fig:Scheme}(a).
The single-electron eigenstates for $A_0=0$ correspond to the Landau levels, with eigenfunctions
$\psi_{n\kappa}(x,y)$ [cf. Eq.~\eqref{Eq:LL} in App.~\ref{App:Details}] and energies $E_{n\kappa}=(n\hbar\oc + 1/2)$, depending on
the cyclotron frequency $\oc=eB/m_{\star}$ and the magnetic length $\lc=\sqrt{\hbar/(eB)}$.
The quantum number $n$ is a non-negative integer, while $\kappa$ is an integer such that $|\kappa|< \AA/(4\pi\lc^2)$. Each level has degeneracy $\NN_{\rm deg} = \AA/(2 \pi \lc^2)$. In the following, we will omit the spin degrees of freedom (we consider magnetic fields where the Zeeman splitting can be neglected).

In the framework of second quantization, we introduce the fermionic operator $\hat{c}^\dagger_{n\kappa}$ ($\hat{c}_{n\kappa}$) which creates (annihilates) an electron in the single-particle state $\ket{n \kappa}$.  
It is convenient to introduce the collective excitation operator
\begin{equation}
\hat{b}^{\dagger} = \frac{1}{\sqrt{\NN_{e}}} 
\sum_{n\kappa}^{n\neq0} \sqrt{n} \, \hat{c}^\dagger_{n\kappa}\hat{c}_{n-1\,\kappa},
\end{equation}
where $\NN_{e}$ is the number of electrons in the 2DEG.
Indeed, after some algebra resumed in App.~\ref{App:SecondQuant}, the current density operator  $\hat{\bf J}=-\frac{e}{m_\star}\sum_i\left(\hat{\bf p}_i+e\hat{\bf A}_i\right)$ can be expressed
in terms of such collective operator, namely:
\begin{eqnarray}\label{Eq:CurrentOperators}
	\hat{J}_x & = &-\sqrt{\frac{\hbar\oc\,\NN_{e} e^2}{2m_\star}}
	\left[\imath\left ( \hat{b} - \hat{b}^{\dagger} \right) + \frac{2\Omega}{\oc} \left(\hat{a}+\hat{a}^\dagger\right)\right], \nonumber\\
	\hat{J}_y & = & -\sqrt{\frac{\hbar\oc\,\NN_{e} e^2}{2m_\star}} \left( \hat{b} + \hat{b}^{\dagger} \right).
\end{eqnarray}

The collective excitation operator $\hat{b}^{\dagger}$ is a `bright' operator, because the light-matter interaction ~\eqref{Eq:2DEGHamiltonianRealSpace} 
can be recast in terms of such operator and the photon operators: 
\begin{equation}
\label{Eq:LightMatterHamiltonian}
\hat{\HH}_{\rm lm}=\hbar\oz\,\hat{a}^\dag \hat{a} +\hat{\HH}_e +\hat{\HH}_I+\hat{\HH}_D,
\end{equation}
where $\hat{\HH}_e = \hbar\oc\,\sum_{n\kappa} n\, \hat{c}_{n\kappa}^\dagger \hat{c}_{n\kappa}$ is the bare electron contribution.
The light-matter coupling $\hat{\HH}_{I}$ reads
\begin{equation}\label{Eq:InteractionHamiltonian}
\hat{\HH}_{I}= \imath\,\hbar\Omega
(\hat{a}+\hat{a}^\dag)(\hat{b} -\hat{b}^{\dagger}),
\end{equation} 
while the diamagnetic energy term $\hat{\HH}_D $ is given by  
\begin{equation}\label{Eq:DiamagneticTerm}
\hat{\HH}_D =
\frac{\hbar\Omega^2}{\oc}\,
(\hat{a}+\hat{a}^\dag)^2.
\end{equation}
Both $\hat{\HH}_{I}$ and $\hat{\HH}_{D}$ depend on $\Omega$, the collective polariton Rabi frequency, defined as
\begin{equation}\label{Eq:DefinitionOmega}
\hbar\,\Omega=e A_0\,\sqrt{\NN_{e}\,\frac{\hbar\oc}{2m_\star}}.
\end{equation}

For non-integer filling factors $\nu = \NN_{e}/\NN_{\rm deg}$, the bare many-body ground state of $\hat{\HH}_e$ is degenerate.
We call $\bar{n}$ the quantum number of the Landau level partially filled by $\NN_{\bar{n}} $ electrons, i.e., $\nu = \bar{n} + \NN_{\bar{n}}/\NN_{\rm deg}$. The generic bare ground state of $\hbar\oz\,\hat{a}^\dag \hat{a} +\hat{\HH}_e$ is
\begin{equation}\label{Eq:FSDefinition}
\ket{{\rm FS},\zeta}=\prod_{n<\bar{n}}\prod_\kappa \hat{c}^\dagger_{n\kappa}\,\prod_{\kappa\in\{\bar{\kappa}\}_\zeta}\hat{c}^\dagger_{\bar{n}\kappa}\,\ket{\rm vac}, 
\end{equation} 
where $\ket{\rm vac}$ is the electron and photon vacuum.
The degeneracy of the Fermi sea is equal to the number of distinguishable permutations for the $\NN_{\bar{n}}$ electrons in the $\NN_{\rm deg}$ possible states.
The set $\{\bar{\kappa}\}_\zeta$ in Eq.~\eqref{Eq:FSDefinition} corresponds to the $\zeta$-th permutation.
For each $\zeta$, we can identify a bright-excitation sector, spanned by the states $\hat{a}^{\dagger m}\hat{b}^{\dagger s} \ket{{\rm FS},\zeta}$ ($m,s\in\mathbb{N}$).
In the thermodynamic limit ($\NN_{e} \gg 1$), the bright excitation operator $\hat{b}^{\dagger}$ behaves as a bosonic operator.
Hence, within the considered bright-excitation sector, $\hat{\HH}_e$ can be replaced by the effective bosonic Hamiltonian
$\hat{\HH}_e \rightarrow E_{\rm FS} + \hbar\oc\,\hat{b}^\dagger\hat{b}$, where $E_{\rm FS}$ is the energy of the Fermi sea.
We emphasize that the light-matter interaction does not couple bright sectors originating from different Fermi seas (i.e., having $\zeta'\neq\zeta$).
Hence, $\hat{\HH}_{\rm lm}$~\eqref{Eq:LightMatterHamiltonian} can be block-diagonalized with one block for each $\zeta$.

\subsection{Linear-response dc conductivity}

To determine the linear response of the 2DEG under the action of a dc electric bias, we follow a Kubo approach \cite{AllenChapter}.
Knowing the manybody eigenstates $\ket{\xi}$ and energies $E_\xi$, the dc magnetoconductivity reads:
\begin{equation}\label{Eq:KuboManyBody}
\sigma_{ij}^{\rm dc} = \imath  \sum_{\xi\neq\xi'} \frac{\emath^{-\beta E_{\xi'}}-\emath^{-\beta E_\xi}}{\AA Z (E_\xi-E_{\xi'}) }
\frac{\braket{\xi|\hat{J}_j|\xi'}\!\braket{\xi'|\hat{J}_i|\xi}}{\left(\omega_\xi-\omega_{\xi'}\right)+\imath/\tau_{\xi\xi'}},
\end{equation}
where $i,j \in \{x,y \}$ and $\beta =1/(k_B T)$ is the inverse thermal energy. Importantly, the current operator $\hat{\bf J}$ in Eq.~\eqref{Eq:CurrentOperators} depends only on the collective bright operators and the photon operators. Hence, in order  to investigate the effects of the light-matter coupling on the magnetoconductivity, we can restrict our treatment to the bright sector, where the Hamiltonian $\hat{\HH}_{\rm lm}$~\eqref{Eq:LightMatterHamiltonian} can be exactly diagonalized through a Hopfield-Bogoliubov transformation \cite{Hopfield,CiutiPRB2005,HagenmullerPRB2010}:
\begin{equation}
\hat{\HH}_{\rm lm} =  E_{\rm GS}+  \hbar \omega_{LP} \,  \hat{p}^{\dagger}_{LP} \hat{p}_{LP} + \hbar \omega_{UP} \,  \hat{p}^{\dagger}_{UP} \hat{p}_{UP},
\end{equation}
where $E_{\rm GS}$ is the  ground-state energy, while  $\omega_{LP}$ ($\omega_{UP}$) is the frequency of the lower (upper) polariton excitation, whose bosonic creation operator is $\hat{p}^{\dagger}_{LP}$ ($\hat{p}^{\dagger}_{UP}$).
Each ground state is now a polariton vacuum, such that $\hat{p}_{LP} \vert {\rm GS,\zeta} \rangle = 0 = \hat{p}_{UP} \vert {\rm GS,\zeta\rangle} $,
and the degeneracy is not changed by the light-matter interaction.
The polariton operators are given by $
\hat{p}_{r} = w_r \hat{a} + x_r \hat{b} + y_r \hat{a}^{\dagger} + z_r \hat{b}^{\dagger} $
with $r \in \{LP,UP \}$.
The vector $\vec{v}_{r} = (w_r, x_r, y_r, z_r)^T$ satisfies the eigenvalue equation $M \vec{v}_{r} = \omega_r \vec{v}_{r}$, where
\begin{equation}
\label{Hopfield}
M =
\begin{pmatrix}
\oz +  2 D& -\imath \Omega & -2 D & -\imath \Omega \\
\imath \Omega & \oc & -\imath \Omega & 0 \\
2 D &   -\imath \Omega & - \oz -  2  D & - \imath \Omega \\
-\imath\Omega & 0  & \imath \Omega &  -\oc
\end{pmatrix},
\end{equation}
being $D = \Omega^2/\oc$ due to the diamagnetic term~\eqref{Eq:DiamagneticTerm}.
The Hopfield-Bogoliubov coefficients satisfy the normalization condition $\vert w_r \vert^2 + \vert x_r \vert^2 - \vert y_r \vert^2 - \vert z_r \vert^2  = 1$.
The anomalous coefficients $y_r$ and $z_r$ are different from zero due to the anti-resonant (non-rotating-wave) terms of the light-matter interaction, which become significant in the ultrastrong coupling regime \cite{CiutiPRB2005}.
The electronic (photonic) weight of the polariton mode $r$ is  $W_{e,r}= \vert x_r \vert^2 - \vert z_r \vert^2$ ($W_{p,r}=\vert w_r \vert^2 - \vert y_r \vert^2$).
It can be shown that $W_{e,r} + W_{p,r} = 1$ and  $\sum_r W_{e,r} = \sum_r W_{p,r} =1$.
A typical polaritonic dispersion \footnote{When the geometric size $L_x$ (and/or $L_y$) is small enough, the Kohn's theorem \cite{Kohn1961} for translationally-invariant systems cannot be applied and Coulomb magnetoplasmon corrections \cite{ParaviciniPRB17} of the cyclotron frequency can occur in the limit of low magnetic field}  is plotted in Fig.~\ref{Fig:Scheme}(b).
As shown in App.~\ref{App:HopfieldBogoliubov}, it is possible to rewrite the current operators~\eqref{Eq:CurrentOperators} in terms of polariton operators as
\begin{align}\label{Eq:CurrentOperatorsNew}
\hat{J}_x=&-\sqrt{\frac{\hbar\oc e^2\NN_{e}}{2m_\star}}
\sum_{r} \frac{\imath\,\omega_r}{\oc}\left[\left(x_r-z_r\right)^*\hat{p}_r-{\rm H.c.}\right],
\nonumber\\
\hat{J}_y=&-\sqrt{\frac{\hbar\oc e^2\NN_{e}}{2m_\star}}
\sum_{r} \left[\left(x_r-z_r\right)^*\hat{p}_r+{\rm H.c.}\right].
\end{align}

\begin{figure}[t!]
	\includegraphics[width=0.95 \linewidth]{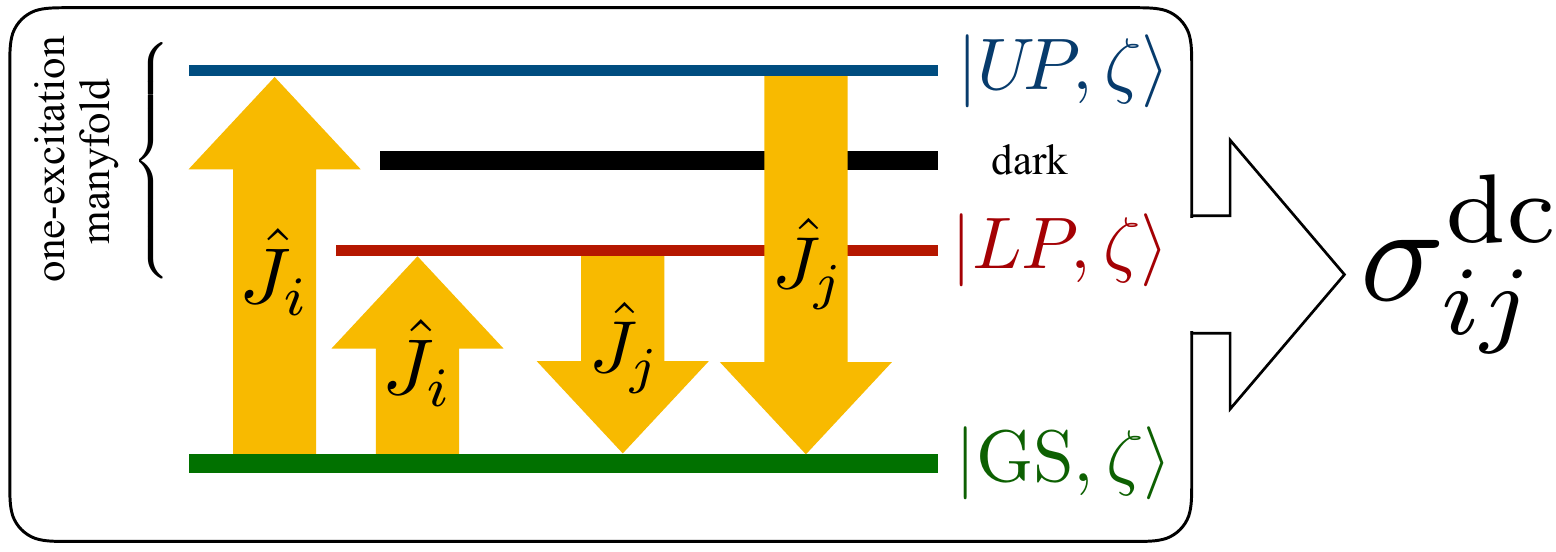}
	\caption{Sketch depicting the role of virtual polaritons in controlling the dc conductivity.
	The low-temperature dc conductivity depends on the matrix elements of the current operator between the ground state and excited states.
		The spatial components of the current operator~\eqref{Eq:CurrentOperatorsNew} couple the manybody ground state $\ket{\rm GS, \zeta}$ only to the one-polariton states $\ket{LP, \zeta}$ and $\ket{UP, \zeta}$.
			The coupling to these virtual excited states determines the $ij$ component of  $\boldsymbol{\sigma}^{\rm dc}$ at $T\to0$.
			The dark states (orthogonal to the bright states) give no contribution to the current matrix elements.
			}
	\label{Fig:Theory}
\end{figure}

In the low-temperature limit ($\beta\to +\infty$), the expression~\eqref{Eq:KuboManyBody} for the magnetoconductivity can be simplified since only for $\ket{\xi}=\ket{{\rm GS},\zeta}$ or $\ket{\xi'}=\ket{{\rm GS},\zeta}$ the contribution to the sum is nonzero. In other words, only matrix elements of the current between the ground state and an excited state matter. Moreover, another remarkable simplification occurs as the only excited states coupled to $\ket{{\rm GS},\zeta}$ by the current operators~\eqref{Eq:CurrentOperatorsNew} are the one-polariton states $\ket{LP, \zeta}= p^{\dagger}_{LP} \ket{{\rm GS}, \zeta}$ and $\ket{UP, \zeta}= p^{\dagger}_{UP} \ket{{\rm GS}, \zeta}$,
as schematically represented in Fig.~\ref{Fig:Theory}. In other words, the dc conductivity depends on the polaritons, which act as virtual excitations.

After some algebra, summarized in App.~\ref{App:Kubo}, it possible to obtain the analytic result of the dc conductivity:
\begin{equation}\label{cond_tensor}
\boldsymbol{\sigma}^{\rm dc}=\frac{n_e e^2}{m_\star}
\sum_{r} \frac{\left|x_r-z_r\right|^2 \tau_r}{1+(\omega_r\tau_r)^2}
\begin{pmatrix}
\frac{\omega_r}{\oc} && -\omega_r\tau_r \\
\omega_r\tau_r && \frac{\oc}{\omega_r}
\end{pmatrix},
\end{equation}
where $n_e = {\NN_{e}}/{\mathcal A}$ is the density of electrons.
The resistivity tensor $\boldsymbol{\rho}^{\rm dc}$ can be obtained by inverting  $\boldsymbol{\sigma}^{\rm dc}$.

The formula~\eqref{cond_tensor} shows that the dc bulk magnetoconductivity tensor of the cavity-embedded 2DEG depends on the cavity-induced change of the ground state (polariton vacuum) and bright excited states (polaritons).
Note that the diagonal components are different, an asymmetry due to the in-plane linear polarization of the cavity mode.
Another crucial ingredient is the transport scattering time $\tau_r$ entering the Kubo conductivity.
This can be written as the sum of two contributions, depending on the electronic and photonic weights as
\begin{equation}\label{Eq:Tau}
\frac{1}{\tau_r}=\frac{W_{e,r}}{\tau_e} + \frac{W_{p,r}}{\tau_p},
\end{equation}
where $\tau_e$ is the electronic transport scattering time (typically due to disorder) and $\tau_p$ is a transport scattering time due to environmental fluctuations affecting the cavity mode (it can be much longer than the cavity photon lifetime). 

Note that for no cavity coupling ($\Omega =  0$), we recover the standard Drude-like magnetoconductivity tensor~\cite{QuantumHallbook}:
\begin{equation}
\boldsymbol{\sigma}^{\rm dc}_{\Omega=0}=
\frac{n_e e^2}{m_\star}
\frac{\tau_{e}}{1+\left(\oc\tau_{e}\right)^2}
\begin{pmatrix}
1 && -\oc\tau_{e} \\ \oc\tau_{ e} && 1
\end{pmatrix},
\end{equation}
and the magnetoresistivity tensor
\begin{equation}\label{Eq:ResistivityNoninteracting}
\boldsymbol{\rho}^{\rm dc}_{\Omega=0}=
\frac{m_\star}{n_e e^2}
\begin{pmatrix}
1/\tau_{e} && \oc \\ -\oc && 1/\tau_{e}
\end{pmatrix}.
\end{equation}

In the regime of Shubnikov-de~Haas (SdH) oscillations and in the low-temperature limit, the electron transport time depends on the single-particle density of states and can be modeled as \cite{Ando1982,IhnBOOK}
\begin{equation}\label{Eq:SdHModel}
\frac{1}{\tau_e} =  \frac{1}{\tau_0}
\left[
1-2\exp\left(-\frac{\pi}{\tau_q\oc}\right)\cos\left(2\pi\nu\right)
\right],
\end{equation}
where $\tau_0$ is the Drude transport time at $B=0$ and $\tau_q$ is the so-called quantum lifetime.

\section{Discussion of results}
\label{results}
\begin{figure}[t!]
	\includegraphics[width=0.9 \linewidth]{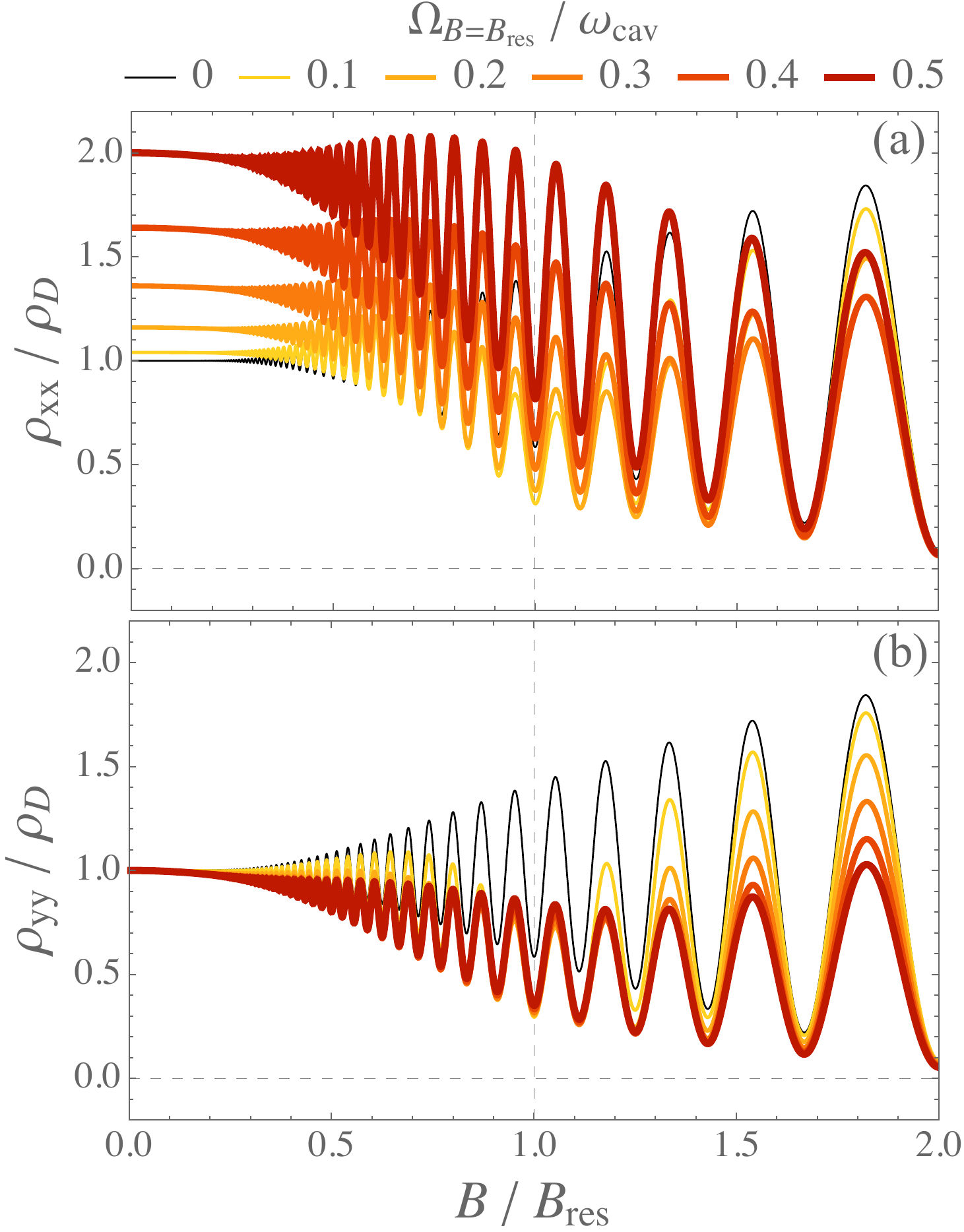}
	\caption{
		Diagonal components of the dc resistivity tensor normalized to the Drude value $\rho_D  = \frac{m_\star  }{n_e e^2\tau_0}$ (no cavity coupling, $B= 0$) \textit{vs} the magnetic field $B$. The different curves correspond to different values of the collective vacuum Rabi frequency at resonance $\Omega_{B=B_{\rm res}}$.
		Thicker lines correspond to larger couplings (cf. legend). The black thin solid line corresponds to the case with no cavity coupling ($\Omega = 0$).
		Top panel: longitudinal resistivity along the cavity mode polarization direction $x$. Bottom panel: longitudinal component along the $y$-direction. 
		Parameters: $\tau_0\oz=100$, $\tau_q\oz=2$, and $\nu_{B=B_{\rm res}}=10$, $\tau_p \gg \tau_e$.
	}\label{Fig:Rho}
\end{figure}

\begin{figure}[t!]
	\includegraphics[width=0.9 \linewidth]{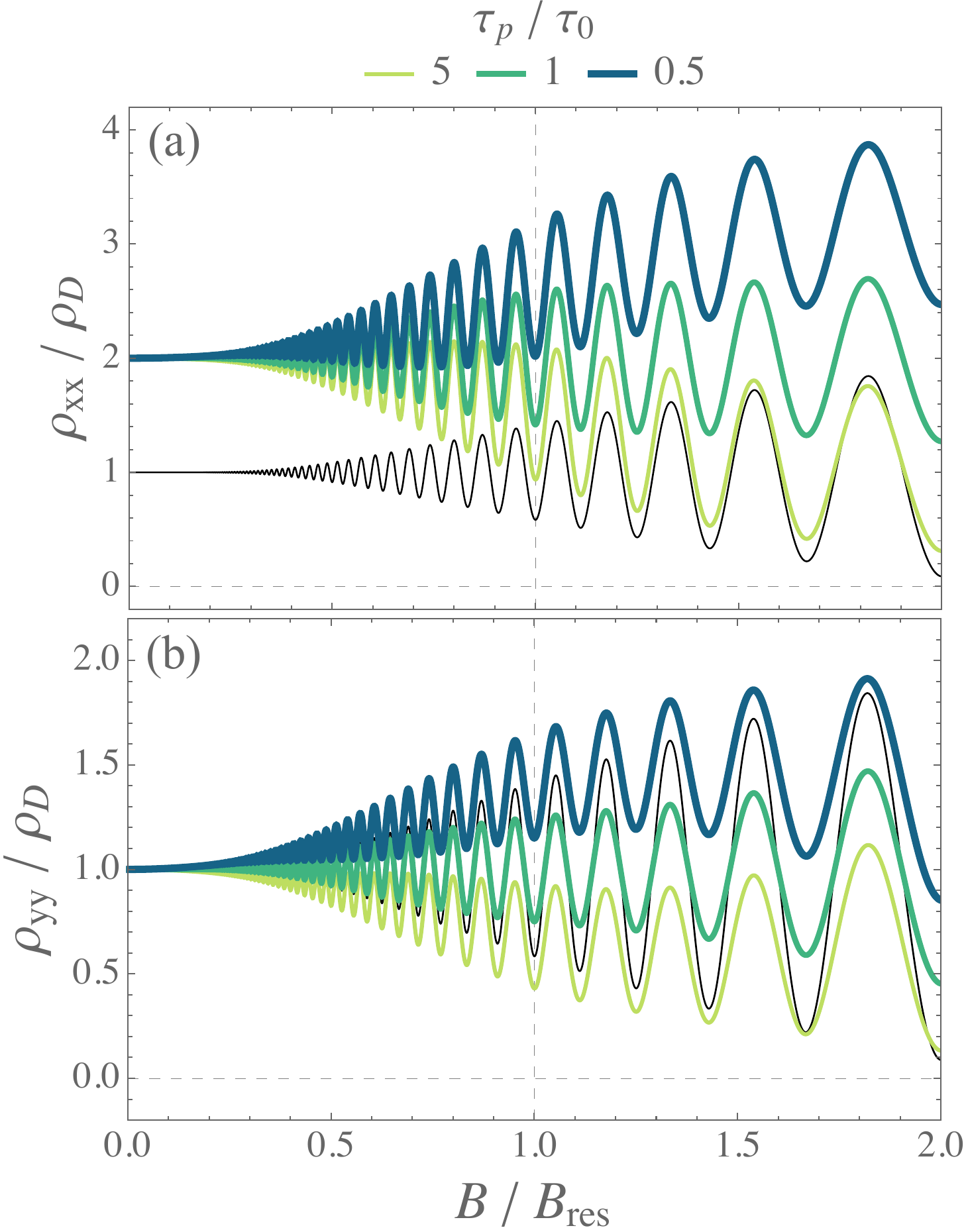}
	\caption{
		Same as Fig.~\ref{Fig:Rho}, but fixing $\Omega_{B=B_{\rm res}}=0.5\,\oz$.
		Different curves (and thicknesses) correspond to different values of $\tau_p/\tau_0$ (cf. legend).
		The black thin solid curve is the reference for no cavity coupling ($\Omega = 0$).
	}\label{Fig:RhoTau}
\end{figure}

In Fig.~\ref{Fig:Rho}, we plot the predictions of our theory for the diagonal components \footnote{The off-diagonal components of the bulk magnetoresistivity (not shown) exhibit negligible cavity-induced changes for the regime of parameters considered in this work} of the resistivity tensor as a function of $B$.
Different curves correspond to different values of $\Omega_{B = B_{\rm res}}$, the collective vacuum Rabi frequency $\Omega$ for $B = B_{\rm res}$
such that $\oc(B_{\rm res}) = \oz$.
Here we consider $\tau_p\gg\tau_e$, i.e., the transport scattering time depends only on the electronic weight of the excitations.
The dc longitudinal resistivity shows typical SdH oscillations, but the envelope is significantly modified by the coupling to the cavity mode.
Panel~\ref{Fig:Rho}(a) displays the results for the diagonal resistivity along the $x$-direction (parallel to the cavity-mode polarization).
The main effect here is an overall increase of the resistivity for decreasing magnetic field.
Indeed, in the limit of low magnetic field, we analytically derived (cf. App.~\ref{App:B0})
\begin{equation}\label{Eq:Rhoxx}
\lim_{B \to 0} \frac{\rho^{\rm dc}_{xx}}{\rho^{\rm dc}_{xx,\Omega=0}} = 1 + 4 \left (\frac{ \Omega_{B = B_{\rm res}}}{\omega_{\rm cav}} \right )^2. 
\end{equation}
Such enhancement becomes quantitatively important in the ultrastrong light-matter coupling regime. For $\Omega_{B = B_{\rm res}} = 0.5\, \oz$, the enhancement is exactly a factor $2$, in agreement with the numerical plot in Fig.~\ref{Fig:Rho}(a). Note that such enhancement is already approached for relatively large magnetic fields $B/B_{\rm res} \sim 0.5$.

Panel~\ref{Fig:Rho}(b) displays the results for the diagonal resistivity perpendicular to the cavity mode polarization vector, which has been measured in recent experiments~\cite{ParaviciniarXiv} \footnote{The experiments in Ref.~\cite{ParaviciniarXiv} have found a behavior of the longitudinal resistance perpendicular to the cavity mode polarization in good agreement with the theory presented here. The longitudinal component parallel to the polarization has not been measured yet, requiring a different design of the cavity-embedded Hall bar}.
Overall, the amplitude of the oscillations is reduced and around $B = B_{\rm res}$ the mean value of the resistivity is suppressed, an effect which is due to the cavity-induced change of the hybrid scattering times $\tau_r$. However, for $B\to0$ we retrieve the $\Omega=0$ behavior~\eqref{Eq:ResistivityNoninteracting}, that is the standard Drude resistivity $\rho_D = \frac{m_\star  }{n_e e^2\tau_0}$ for $B=0$ and no cavity.

In Fig.~\ref{Fig:RhoTau}, we present our predictions for finite $\tau_p$, taking  $\Omega_{B=B_{\rm res}} = 0.5\, \oz$.
When $\tau_p>\tau_0$,  the phenomenology is similar to Fig.~\ref{Fig:Rho} ($\tau_p \gg \tau_0$). When $\tau_p = \tau_0$, the SdH oscillations become symmetric with respect to their mean value.
For $\tau_p < \tau_0$ and relatively large $B$ also the resistivity $\rho^{\rm dc}_{yy}$ is increased.  In the limit of low magnetic fields, however, the phenomenology is robust with respect to the ratio $\tau_p/\tau_0$ for both longitudinal components of $\boldsymbol {\rho}^{\rm dc}$.

\section{Conclusions} \label{Sec:Conclusions}

In conclusion, we have derived an analytical theory showing how the bulk magnetotransport of a 2DEG can be strongly modified by the coupling to a cavity photon mode with in-plane linear polarization. The results are remarkable since strong modifications and anisotropy appear in the dc linear resistivity in a regime where no real photons are injected nor created. An intriguing perspective is the study of the quantum Hall regime when the bulk is insulating and the transport is due to the edge states.
From a general point of view, our work shows that the dc transport of cavity-embedded electronic systems is controlled by virtual polariton excitations and that the electronic properties of materials can be dramatically controlled by the vacuum field of electromagnetic cavity resonators.

\acknowledgements
We would like to thank J. Faist, G. L. Paravicini-Bagliani and G. Scalari for discussions and for showing experimental results prior to publication.
We are grateful to A. Biella and I. Carusotto for a critical reading of the manuscript.

\appendix
\section{Details about the derivation}\label{App:Details}

\subsection{Many-body operators in second-quantization}\label{App:SecondQuant}

In order to cast the Hamiltonian $\hat{\HH}_{\rm lm}$ and current operator $\hat{\bf J}$ in second-quantized form, we have exploited the orthogonality relations between the Landau levels eigenfunctions
\begin{align}\label{Eq:LL}
\psi_{n\kappa}(x,y) & =  \left(\sqrt{\pi}\, 2^n n!\, L_y \lc\right)^{-1/2}
{\rm H}_n\!\left(\!\frac{x-x_\kappa}{\lc}\!\right)
\nonumber\\
& \exp\left[-\frac{\left(x-x_\kappa\right)^2+2\imath x_\kappa y}{2\lc^2}\right],
\end{align}
where ${\rm H}_n$ represents the $n$-th order Hermite polynomial and $x_\kappa=-2\pi\kappa\lc^2/L_y$ is the orbit center position.
Hence, we have
\begin{align}\label{Eq:SecondQuant}
&\sum_i \hat{p}_{x,i} = \imath \frac{\hbar}{\lc} \sqrt{\frac{\NN_{e}}{2}} \left(\hat{b} - \hat{b}^\dagger \right),
\nonumber\\
&\sum_i \left(\hat{p}_{x,i} + eB \hat{x}_i\right) = \frac{\hbar}{\lc} \sqrt{\frac{\NN_{e}}{2}} \left(\hat{b} + \hat{b}^\dagger \right),
\end{align}
from which one easily gets the form of $\hat{J}_x$, $\hat{J}_y$, $\hat{\HH}_I$ given in Eqs.~\eqref{Eq:CurrentOperators} and~\eqref{Eq:DiamagneticTerm}.

\subsection{On the Hopfield-Bogoliubov coefficients}\label{App:HopfieldBogoliubov}

To rewrite the current operators~\eqref{Eq:CurrentOperators} in the form~\eqref{Eq:CurrentOperatorsNew}, we first express $\hat{b}$ and $\hat{a}$ in terms of the polariton operators $\hat{p}_r$ via
\begin{align}
\hat{a}&=\sum_r \left(w^*_r\, \hat{p}_r - y_r\, \hat{p}_r^\dagger \right),
\nonumber\\
\hat{b}&=\sum_r \left(x^*_r\, \hat{p}_r - z_r\, \hat{p}_r^\dagger \right).
\end{align}
Furthermore, we consider that, by definition, the eigenvector $\vec{v}_{r} = (w_r, x_r, y_r, z_r)^T$ satisfies $\vec{v}_{r} = \omega_r M^{-1} \vec{v}_{r}$. Hence, inverting explicitly the matrix $M$ given in Eq.~\eqref{Hopfield}, we found the exact relation
\begin{equation}
\oc(x_r + z_r) +2 \imath\Omega(w_r-y_r) = \omega_r(x_r-z_r),
\end{equation}
which allows us to write both $\hat{J}_x$ and $\hat{J}_y$ in terms of the electronic coefficients $x_r$ and $z_r$ only.
With a similar procedure, it is possible to obtain another useful relation, namely
\begin{equation}\label{Eq:App_Wr}
\left|x_r-z_r\right|^2=W_{e,r} \frac{\omega_r}{\oc}.
\end{equation}

\subsection{On the linear dc conductivity}\label{App:Kubo}

Let us start from Eq.~\eqref{Eq:KuboManyBody} and consider the low-temperature limit $\beta\to+\infty$.
As discussed above, in this regime only the terms with $\ket{\xi},\ket{\xi'}=\ket{{\rm GS}, \zeta}$ give a nonzero contribution to the sum, the others being suppressed by the Boltzmann coefficients.
After a simple substitution, Eq.~\eqref{Eq:KuboManyBody} can be cast in the form
\begin{widetext}
\begin{equation}\label{Eq:SigmaIntermediate}
\sigma_{ij}^{\rm dc}=\imath \frac{\emath^{-\beta E_{\rm GS}}}{\AA Z} \sum_{\zeta=1}^{\mathcal{D}_{\rm GS}} \sum_{r}
\frac{1}{\hbar \omega_r}
\left(
\frac{\braket{r,\zeta|\hat{J}_j|{\rm GS},\zeta}\!\braket{{\rm GS},\zeta|\hat{J}_i|r,\zeta}}
{\omega_r+\imath/\tau_r}
+
\frac{\braket{{\rm GS},\zeta|\hat{J}_j|r,\zeta}\!\braket{r,\zeta|\hat{J}_i|{\rm GS},\zeta}}
{-\omega_r+\imath/\tau_r}
\right),
\end{equation}
\end{widetext}
where we used the fact that the current operators~\eqref{Eq:CurrentOperatorsNew} can only couple $\ket{{\rm GS}, \zeta}$ to the excited states $\ket{r, \zeta}$.
In Eq.~\eqref{Eq:SigmaIntermediate}, $\tau_r$ is the transport scattering time associated to the polaritonic transition [Eq.~\eqref{Eq:Tau}], while $\mathcal{D}_{\rm GS}$ is the degeneracy of the manybody ground state, which is unaltered with respect to that of the Fermi sea $\mathcal{D}_{\rm FS}$.
Due to the form of the $\hat{J}_i$ operators, each $\zeta$ gives the same contribution to $\sigma_{ij}^{\rm dc}$.
In the zero-temperature limit, this multiplicity is regularized by the partition function, which simply becomes $Z\xrightarrow{\beta\to+\infty}\mathcal{D}_{\rm GS}\emath^{-\beta E_{\rm GS}}$.
This allows to further simplify Eq.~\eqref{Eq:SigmaIntermediate} as
\begin{align}
\sigma_{ij}^{\rm dc}
&=\frac{1}{\AA} \sum_r
\frac{\tau_r}{\hbar \omega_r}
\left(
\frac{\Theta_{ij}^{(r)}}
{1-\imath \omega_r\tau_r}
+
{\rm c.c.}
\right)
\nonumber\\
&=\frac{1}{\AA} \sum_r
\frac{2\tau_r}{\hbar \omega_r}
\frac{\Re{\Theta_{ij}^{(r)}}-\omega_r\tau_r\,\Im{\Theta_{ij}^{(r)}}}
{1+\left(\omega_r\tau_r\right)^2},
\end{align}
where we have introduced the quantities
\begin{equation}
\Theta_{ij}^{(r)}=\braket{{\rm GS},\zeta|\hat{J}_i|r,\zeta}\!\braket{r,\zeta|\hat{J}_j|{\rm GS},\zeta},
\end{equation}
which do not depend on $\zeta$ and can be calculated exactly via Eqs.~\eqref{Eq:CurrentOperatorsNew}.
Some straightforward algebra finally gives the form presented in Eq.~\eqref{cond_tensor}.

\subsection{Limit $B\to0$}\label{App:B0}

To obtain the analytic result~\eqref{Eq:Rhoxx} for $B\to0$, we first consider that, using relation~\eqref{Eq:App_Wr}, the conductivity tensor~\eqref{cond_tensor} can be cast as
\begin{equation}
\boldsymbol{\sigma}^{\rm dc}=\frac{n_e e^2}{m_\star}
\sum_{r} \frac{W_{e,r}}{\oc} \frac{\omega_r \tau_r}{1+(\omega_r\tau_r)^2}
\begin{pmatrix}
\frac{\omega_r}{\oc} && -\omega_r\tau_r \\
\omega_r\tau_r && \frac{\oc}{\omega_r}
\end{pmatrix}.
\end{equation}
In the $B\to0$ limit, the factor $W_{e,r}$ in Eq.~\eqref{cond_tensor} will select only the contribution coming from the lower polariton.
Correspondingly, for $\tau_p\gg\tau_0$, Eqs.~\eqref{Eq:Tau} and~\eqref{Eq:SdHModel} imply $\tau_{LP}\to\tau_0$.
In this regime, the conductivity tensor~\eqref{cond_tensor} can be easily inverted to give
\begin{equation}\label{Eq:ResistivityB0}
\boldsymbol{\rho}^{\rm dc}\xrightarrow{B\to 0}
\frac{m_\star}{n_e e^2 \tau_0}
\begin{pmatrix}
\left(\frac{\oc}{\omega_{LP}}\right)^2 && \oc\tau_{0} \\ -\oc\tau_{0} && 1
\end{pmatrix}.
\end{equation}
The off-diagonal components and the diagonal one for the direction perpendicular to the field polarization are unchanged with respect to the noninteracting case~\eqref{Eq:ResistivityNoninteracting}. On the contrary, the component along the field polarization depends on the ratio $\omega_{LP}/\oc$.
As it can be seen in Fig.~\ref{Fig:Scheme}(b), due to the light-matter interaction the slope at which $\omega_{LP}$ goes to zero is different than that for $\oc$.
More specifically, one has analytically
\begin{equation}
\left(\frac{\omega_{LP}}{\oc}\right)^2 = \frac{\oz^2}{\oz^2+4\Omega_{B = B_{\rm res}}^2} + \mathcal{O}\left[B^2\right],
\end{equation}
from which Eq.~\eqref{Eq:Rhoxx} follows.
This change in slope of $\omega_{LP}$ is the counterpart of the opening of the so-called polariton gap between the two polariton branches~\cite{HagenmullerPRB2010,ScalariScience2012}.

\bibliographystyle{apsrev4-1}
\bibliography{bibliography}

\end{document}